\documentclass[11pt]{article}
\usepackage{amsmath}
\usepackage{amsgen,amstext,amsbsy,amsopn,amsthm, amssymb}
\usepackage{amsmath,amsfonts,amsthm,amssymb}

\newcommand\version{August 29th, 2008}
\numberwithin{equation}{section}

\newtheorem{mydef}{DEFINITION}

\newcommand{\beq}{\begin{equation}}
\newcommand{\eeq}{\end{equation}}

\newcommand{\beqa}{\begin{eqnarray}}
\newcommand{\eeqa}{\end{eqnarray}}

\newcommand\al{\alpha}

\newcommand\N{{\mathbb N}}

\newcommand\R{{\mathbb R}}
\newcommand\eps{\varepsilon}
\newcommand\half{\mbox{$\frac 12$}}

\newcommand\rhov\varrho

\newcommand\const{{\rm const.\, }}

\newcommand\RR{{\widetilde R}}

\newcommand\TR{\theta_{\widetilde R}}
\newcommand\BTR{\bar\theta_{\widetilde R}}

\renewcommand\aleph\varaleph
\renewcommand\rho\varrho

\newtheorem{thm}{THEOREM}

\newtheorem{lem}{Lemma}
\newtheorem{cor}{COROLLARY}

\begin{document}

\markboth{\scriptsize{\version}}{\scriptsize{\version}}
\title{\bf{The Ground State Energy of Dilute Bose Gas in Potentials with Positive Scattering Length}}
\author{\vspace{5pt} Jun Yin 
\\ \vspace{-2pt}\small{ Department of Mathematics, Harvard
University,  }\\ \vspace{-2pt}\small 
\\ {\small Email: \texttt  {JYin@math.harvard.edu}} }

\date{\small \version}
\maketitle

\begin{abstract}
The leading term of the ground state energy/particle  of a dilute gas of bosons with mass $m$ in the thermodynamic limit is $2\pi \hbar^2 a \rho/m$ when the density of the gas is $\rho$, the interaction potential is non-negative and the scattering length $a$ is positive. In this paper, we generalize the upper bound part of this result to any interaction potential with positive scattering length, i.e, $a>0$ and the lower bound part to some interaction potentials with shallow and/or narrow negative parts. 
 \end{abstract}
\renewcommand{\thefootnote}{${\,}$}
\footnotetext{***}
\renewcommand{\thefootnote}{${\, }$}
\footnotetext{\copyright\,2008 by the author.
This paper may be reproduced, in its entirety, for non-commercial
purposes.}

\section{Introduction and main theorems}
In Dyson's work \cite{dyson} and  Lieb, Yngvason and Seiringer's work \cite{LY1, LSY00}, it is rigorously proved that the leading term of the ground state energy/particle of a three dimensional dilute bose gas of mass $m$ in the thermodynamic limit with density $\rho$ is $2\pi \hbar^2 a \rho/m$,  i.e., 
\beq\label{oldmain}
e(\rho, m)=2\pi \hbar^2 a \rho/m (1+o(1)) \,\,\,{\rm if}\,\,\, a^3\rho\ll1 
\eeq
where they assumed that the interaction potential is non-negative, the scattering length $a$ is positive. This result is  generalized to a two dimensional dilute bose gas in \cite{LY2}. In this paper, first, in Theorem 1,  we generalize the upper bound part of (\ref{oldmain}) to  general interaction potentials $v$ with positive scattering length. On the other hand, for the lower bound on the ground energy, it was conjectured in \cite{LY1} that the lower bound part of (\ref{oldmain}) should hold if the scattering length is positive and $v$ has no $N$-body bound states for any $N$.  Recently, it is proved in \cite{JOL} that   in some cases with partly shallow negative potential the lower bound part of (\ref{oldmain}) holds. In Theorem 2, we introduce a different method for the lower bound on (\ref{oldmain}) when $v$ can have shallow and/or narrow  negative components and provide better(smaller) error term. 
\par 
We begin with describing the questions more precisely. We write the Hamiltonian of a system of $N$ interacting bosons which are restricted to a cubic box of volume $\Lambda=L^3$ in the following way (in units  where $\hbar=2m=1$): 
\beqa\label{mainHam}
H_N\equiv\sum_{i=1}^N-\Delta_i+\sum_{1\leq i<j\leq N}v^a(x_i-x_j) 
\eeqa
Here $\Delta$ denotes the Laplacian on $\Lambda$ with periodic boundary condition and $v^a$ is a scaled interaction potential, i.e.,
\beq\label{13}
v^a(x)={a^{-2}}\cdot v(x/a),\,\,\,a>0
\eeq
The pair interaction potential $v$ is spherically symmetric and supported on the set $\{x\in \R^3:|x|
\leq R_0\}$ for some $R_0>0$. 
\begin{mydef}[Scattering Length]
\par Assume that $w$ is a pair spherically symmetric interaction potential with compact support. Denote $E[\phi\,]$ as the energy of the complex-valued function $\phi$ on $\R^3$ as follows,
\beq
E[\phi\,]=\int_{\R^3}|\nabla\phi(x)|^2+\half w(x)|\phi(x)|^2dx.
\eeq
Define the scattering length $SL(w)$ of potential $w$ as the following minimum energy.
\beq
SL(w)\equiv\min_{\phi}\left\{\frac{1}{4\pi}E[\phi\,]: \lim_{|x|\to\infty}\phi(x)=1 \right\}
\eeq
\end{mydef}
Note: If $SL(w)>-\infty$, one can easily prove that the Hamiltonian $-\Delta+\half w$ has no bound state. In particular, when $w\geq 0$, we have $SL(w)\geq 0$ and $w$ has no bound state. One can see that this definition is equivalent to the definition of scattering length in \cite{ERJJ2} when $w>0$. 
\par With the relation between $v$ and $v^a$ in (\ref{13}), we can assume that 
 \beq \label{16}SL(v)=1,\,\,\,\,SL(v^a)=a
 \eeq
\par Let $f_1(x)$ be  the solution of the zero-energy scattering equation of $v$, i.e., 
\beq\label{v0s}
-\Delta f_1(x)+\half v(x)f_1(x)=0,
\eeq
then we have that $f_a(x)\equiv f_1(x/a)$ is the solution of the following zero-energy scattering equation of $v^a$. 
\beq
\label{fa0}-\Delta f_a(x)+\half v^a(x)f_a(x)=0,\eeq
As in \cite{ERJJ2}, one can prove that if $f_a$ is normalized as $\lim_{|x|\to \infty}f_a(x)=1$, then 
\beq\label{pfa1}
f_a(x)=1-a/x,\,\,\,{\rm for}\,\,\,|x|>R_0a
\eeq
In this paper, we are interested in the ground energy $E(N,\Lambda)$ of $H_N$ in the thermodynamic limit that $\Lambda\to \infty$, $N\to\infty$ and $N/\Lambda=\rho$. Low density means that the average inter-particle distance $\rho^{-1/3}$ is much larger than the scattering length $a$, i.e. $a^3\rho\ll 1$.  
\par First, we state that for any fixed $v$, the upper bound on (\ref{oldmain}) holds for the dilute bose gas. 
\begin{thm} Fix $v$ with $SL[v]=1$ and $v^a$ satisfying (\ref{13}). Let $f_1$ be the solution of zero-energy equation of $v$ and normalized as: $f_1(\infty)=1$. In the thermodynamic limit, $\lim_{N\to\infty}N/\Lambda=\rho$, we have the following upper bound on $E(N,\Lambda)$, which is the ground energy of $H_N$ in (\ref{mainHam}),
\beqa\label{mainupper}
\limsup_{N\to\infty}\frac{E(N,\Lambda)}{4\pi a\rho N}\leq 1+\const(a^3\rho)^{1/4},
\eeqa
for some constant depending on $\|f_1\|_{\infty}$, provided that $\frac{4\pi}{3}a^3\rho\leq 1$.
\end{thm}
Note: So far, the best proof of the error term on upper bound, when $v\geq 0$, is $O(a^3\rho)^{1/3}$, as in \cite{ERJJ2}. 
\par On the other hand, for the lower bound in (\ref{oldmain}), we prove that as long as $v$ has a positive core and is bounded from below,  (\ref{oldmain}) holds when the negative part is small enough (shallow and/or narrow). In the appendix, we show that if $v^a$ is a continuous function on $ \R^3$ and $H_N$ has no bound state for any $N$, $v^a$ satisfies the above two requirements, i.e.,  
\beq\label{1.11} v^a(0)>0,\,\,\,\,\min v^a(r)>-\infty\eeq 
The above two inequalities (\ref{1.11}) also hold when $v^a$ is stable \cite{DavidR} (the stability of potential is assumed in \cite{JOL}).
\begin{thm}We assume that $v(x)=v_+(x)+ v_-(x)$, $v_+(x)\geq 0$, $v_-(x)\geq- \lambda_-$, $\lambda_->0$ and $v_+$ has a positive core, i.e. $\exists\, r_1$, such that $v_+(x)\geq \lambda_+>0$ for $|x|\leq r_1$. Here $v_-$ need not be negative. 
\par There exist $ c_1(R_0/r_1)$ and $c_2(R_0/r_1)$, which are greater than one and  only depend on $R_0/r_1$, such that the following holds. 
\par If there exists some positive number $t$ satisfying
\beq SL[c_1(R_0/r_1)\cdot(v+tv_-)]\geq 0\,\,\,{\rm and }\,\,\,\lambda_+\geq (1+t^{-1})\,c_2(R_0/r_1)\cdot\lambda_-,\eeq 
we have  the following lower bound on $E(N,\Lambda)$, 
\beqa
\liminf_{N\to\infty}\frac{E(N,\Lambda)}{4\pi a\rho N}\geq 1-\const(a^3\rho)^{1/17},
\eeqa
for some constant depending on $v_+$ and $v_-$, provided that $\frac{4\pi}{3}a^3\rho$ is smaller than some constant depending on $v_+$, $v_-$ and $t$.
\end{thm}
Note:  So far, the best estimation of the error term of the lower bound, when $v>0$, is also $O(a^3\rho)^{1/17}$, as in \cite{ERJJ2}. 
\par This theorem implies the following two corollaries.
\begin{cor}
Assume that \beq v(x)=v_+(x)+\lambda_- v_-(x),\,\,\, v_+(x)\geq 0, \,\,\,v_-(x)\geq -1\eeq and $v_+$ has a positive core, i.e. $\exists r_1$ such that $v_+(x)\geq \lambda_+$ for $|x|\leq r_1$. There exists $\lambda_0(r_1,R_0,\lambda_+,v_-)$ such that, if $0\leq \lambda_-\leq \lambda_0$, i.e., the potential is shallow enough, we have  the following lower bound on $E(N,\Lambda)$, 
\beqa
\liminf_{N\to\infty}\frac{E(N,\Lambda)}{4\pi a\rho N}\geq 1-\const(a^3\rho)^{1/17}
\eeqa 
provided that $\frac{4\pi}{3}a^3\rho$ is smaller than some constant depending on $v_+$ and $v_-$.
\end{cor} 
\begin{proof}
For fixed $R_0$, $r_1$ and $\lambda_+$, when $\lambda_-$ is small enough, we have that   
\beq
SL[c_1(R_0/r_1)(v_++2\lambda_-v_-)]\geq 0\,\,\,{\rm and }\,\,\,\lambda_+\geq 2\, c_2(R_0/r_1)\lambda_-.
\eeq
 Using Theorem 2, with the choice $t=1$, we arrive at the desired result.
\end{proof}
\begin{cor}
Assume that 
\beq
 v(x)=v_+(x)+ v_-(x),\,\,\,\,v_+(x)\geq 0\geq v_-(x)\geq -\lambda_-
\eeq and $v_+$ has a positive core, i.e. $\exists r_1$ such that  $v_+(x)\geq \lambda_+$ for $|x|\leq r_1$. There exist $\lambda_0(R_0/r_1,\lambda_-)$ and $\eps\,(R_0, r_1,\lambda_-)$ such that, if 
\beq\label{coleq1}
\lambda_+\geq \lambda_0\,\,\,{\rm and }\,\,\,
\int_{x\in \R^3}|v_-(x)|dx\leq \eps\,(R_0, r_1,\lambda_-),
\eeq 
we have  the following lower bound on $E(N,\Lambda)$, 
\beqa\label{coleq2}
\liminf_{N\to\infty}\frac{E(N,\Lambda)}{4\pi a\rho N}\geq 1-\const(a^3\rho)^{1/17}
\eeqa 
provided that $\frac{4\pi}{3}a^3\rho$ is smaller than some constant depending on $v_+$ and $v_-$.
\end{cor} 
\begin{proof}  We choose $\lambda_0=\max\{3,2\,c_2(R_0/r_1)\}\lambda_-$, then we have that $\lambda_+\geq \lambda_0\geq 3\lambda_-$, which implies that
\beq\label{vv-geq}
[v+v_-](x)\geq\lambda_+/3\geq 0\,\,\,{\rm for }\,\,\, |x|\leq r_1
\eeq
  Then we claim that for any $n\geq 1$ and $\lambda_+\geq 3\lambda_-$, there exists $\xi(n)>0$, 
\beq\label{slnvv-}
\int_{ \R^3}|v_-(x)|dx\leq \xi(n)\Rightarrow SL[n(v+v_-)]\geq 0.
\eeq
To prove (\ref{slnvv-}), we shall prove that there exists $\xi(n)>0$, if $\int_{ \R^3}|v_-(x)|dx\leq \xi(n)$, for any  non-negative radial function $f$,
\beq\label{slfg0}
 \int_{|x|\leq R_0}|\nabla f|^2(x)+\frac n2(v+v_-)f^2(x)dx\geq 0.
 \eeq
We can see, with (\ref{vv-geq}),
\beqa\label{120}
&&\int_{|x|\leq R_0}\left[|\nabla f|^2(x)+\frac n2(v+v_-)f^2(x)\right]dx\\\nonumber
\geq && \int_{r_1\leq |x|\leq R_0}\left[|\nabla f|^2(x)-\|nv_-\|_\infty f^2(x)\right]dx\\\nonumber
\geq && \int_{r_1\leq |x|\leq R_0}|\nabla f|^2(x)dx-n\lambda_-\int_{r_1\leq |x|\leq R_0}| f|^2(x)dx
\eeqa
Hence, if (\ref{slfg0}) does not hold, the right side of (\ref{120}) is less than 0.  With Sobolev inequality and Schwarz's Inequality, we obtain that there exists $ \eta(n)$  such that 
\beq\label{123}
\left(\int_{r_1\leq |x|\leq R_0}|f(x)|^4dx\right)^{1/2}\leq \eta(n)\int_{r_1\leq |x|\leq R_0}| f|^2(x)dx
\eeq
On the other hand, with (\ref{vv-geq}), $v+v_-\geq 2v_-$ and Schwarz's Inequality, we have that
\beqa\label{121}
&&\int_{|x|\leq R_0}|\nabla f|^2(x)+\frac n2(v+v_-)f^2(x)dx\\\nonumber
\geq &&\int_{|x|\leq R_0}|\nabla f|^2(x)+\int_{|x|\leq r_1}\frac{n\lambda_+}{6}f^2(x)-\left|\int_{r_1\leq |x|\leq R_0}\!\!\!\!\!\!\!\!\!\!nv_-(x)|f(x)|^2dx\right|\\\nonumber
\geq &&\int_{|x|\leq R_0}|\nabla f|^2(x)+\int_{|x|\leq r_1}\frac{n\lambda_+}{6}f^2(x)
\\\nonumber &&-n\eta(n)\left|\int_{r_1\leq |x|\leq R_0}v^2_-(x)dx\right|^{1/2}\int_{r_1\leq |x|\leq R_0}| f|^2(x)dx
\\\nonumber\geq &&\int_{|x|\leq R_0}|\nabla f|^2(x)+\int_{|x|\leq r_1}\frac{n\lambda_+}{6}f^2(x)
-n\eta(n)\lambda_-\|v_-\|_1^{1/2}\int_{ |x|\leq R_0}| f|^2(x)dx\\\nonumber
\eeqa
Thus, for $n\geq 1$, if
\beqa\nonumber
 &&\|v_-\|_1^{1/2}\leq \left(\xi(n)\right)^{1/2}\equiv \frac{1}{n\eta(n)}\cdot\min_{f}\frac{\int_{|x|\leq R_0}|\nabla f|^2(x)+\int_{|x|\leq r_1}\frac{n\lambda_+}{6}f^2(x)dx}{\int_{|r|\leq R_0}| f|^2(x)dx}, 
 \eeqa
the inequality (\ref{slfg0}) holds. We note that it is easy to see that $\xi(n)>0$. Hence we arrive at the desired result (\ref{slnvv-}). 
At last, choosing 
 \beq
\eps\,(R_0, r_1,\lambda_-)=\xi\left(c_1(R_0/r_1)\right)
 \eeq
  and using the result of Theorem 2 with $t=1$, we arrive at  the desired result (\ref{coleq2}). 
\end{proof}
Remark: Compared with the result of \cite{JOL}, we improve the error term (It was $(a^3\rho)^{1/31}$ in \cite{JOL}) and generalize the shapes of potentials, i.e., the negative part of potential can be shallow and/or narrow. In particular, there is no restriction on the depth of the interaction potential $v$, i.e. for $\forall \lambda_->0$, there $\exists v$ satisfying $\min_{x\in\R^3}v(x)<-\lambda_-$ and Theorem 2 holds. 
\section{Proofs}
\subsection {Proof of Theorem one}
\par
\begin{proof}
As usual, to prove the upper bound on the  ground state energy, we only need to construct a sequence of trial states $\Psi_{N,\Lambda}$ satisfying 
\beq\label{psiny}
\limsup_{N\to\infty}\frac{\langle \Psi|H_N|\Psi\rangle}{N\langle \Psi|\Psi\rangle}\leq 4\pi a\rho  (1+\const Y)
\eeq
for some constant that depends only on $\|f_1\|_\infty$. Here we denote $Y$ as 
\beq
Y\equiv\left(\frac{4\pi}{3}a^3\rho\right)^{1/4}
\eeq 
Following the ideas in \cite{dyson, LSY00}, we construct the trial state of  the following form,
\beq
\Psi_N=\prod_{p=1}^NF_p
\eeq
In \cite{dyson}, $F_p$ depends on the the nearest particle to the $x_p$ among all the $x_i$ with $i<p$, i.e., 
\beq
F_p=f(t_p), \,\,\,t_p=\min_{i<p}\left\{\left|x_i-x_p\right|\right\}
\eeq
via the function $f$ which is very close to the zero energy scattering solution and  satisfies 
\beq
0\leq f\leq 1,\,\,\,\, f\,'\geq 0
\eeq
Hence in \cite{dyson}, $F_p$ has the following property
\beq\label{LYresult}
F_{p,i}\cdot f(|x_p-x_i|)\leq F_p\leq F_{p,i}
\eeq
Here $F_{p,i}$ is defined in \cite{dyson} as  the value that $F_p$ would take if the point $x_i$ were omitted from consideration. 
\par 
But in our case where the potential has a negative part, the zero energy scattering solution $f_a$ of $v^a$ may not be   an increasing function or bounded by 1 (if it was, the proof would be much simpler). Hence we do \textit{not} have the property (\ref{LYresult}). For this reason, our choice of $F_p$ will be more complicated. Our $F_p$ depends on all \textit{particles} near the $x_p$, not just the nearest. 
\par We remark that the function $F_p$ should have following properties.
\begin{enumerate}
	\item $F_p$ is a continuous function of $x_i$ ($1\leq i\leq N$).
	\item When $|x_i-x_p|$ is large enough, the position of $x_i$ does not effect $F_p$, i.e., $\nabla_{x_i}F_p=0$.
	\item $F_p$ has a similar property as (\ref{LYresult}).
\end{enumerate}
\par First we define $\theta_r(x)$  as the characteristic function of the set $\{x : |x| \leq r\}$ and  $\bar\theta_r\equiv1-\theta_r$. Choosing $b=a /Y$, we have 
\beq
 a/b=\frac{4\pi}{3}b^3N/\Lambda=Y.
\eeq
Without loss of generality, we assume that $b> \max\{2R_0a,4a\}$, as in \cite{dyson, ERJJ2}
. We define $f(x)$ as 
\begin{equation}\label{deff}
f(x) = \left\{ \begin{array}{ll} f_a(x)/f_a(b) & b\geq |x|\geq 0 \\ 
1 & {\rm otherwise}\,,
\end{array}\right.
\end{equation}
Here $f_a$ is the zero energy scattering solution of $v^a$, as in (\ref{fa0}). With the equation (\ref{pfa1}), we note that 
\beq\label{prof}
f(x)=\frac{1-a/|x|}{1-a/b},\,\,{\rm for}\,\,\,b\geq|x|\geq R_0a.
\eeq
Let $\RR=\max\{R_0a,2a\}$, which implies that $f(\RR)>\half$. We define 
$\Theta^{in}_p$, $\Theta^{out}_p$  ($1<p\leq N$) as 
\beqa\label{definout}
\Theta^{in}_p\equiv\prod_{j<p}\TR(x_j-x_p),\,\,\,\Theta^{out}_p\equiv\prod_{j<p}\BTR(x_j-x_p)
\eeqa
We can see that $\Theta^{in}_p=1$ when $|x_j-x_p|\leq \RR$ for all $j<p$ and $\Theta^{out}_p=1$ when  $|x_j-x_p|> \RR$ for all $j<p$.
With $\Theta^{in}_p$ and $\Theta^{out}_p$, we can define $r_p(x_1,\cdots,x_N)$ and  $R_p(x_1,\cdots,x_N)$ as follows, ($x_i\in[0,L]^3, i=1,\cdots N$)
\beqa
&&r_p\equiv\\\nonumber
&&(1-\Theta^{out}_p)\cdot \min_{i<p} \bigg\{|x_i-x_p|: f(x_i-x_p)=\min_{j<p} \big\{f(x_j-x_p): |x_j-x_p|\leq\RR\big\}\bigg\}\\\nonumber
&&R_p\equiv\RR\cdot\Theta^{in}_p+(1-\Theta^{in}_p)\times \min_{i<p}\bigg\{|x_i-x_p|: |x_i-x_p|>  \RR\bigg\}
\eeqa
With the definition of $R_p$ and (\ref{prof}), we have that
\begin{enumerate}
	\item $f(R_p)\leq f(x_j-x_p)$ for any $j<p$ satisfying $|x_j-x_p|>\RR$,
	\item $R_p \leq |x_j-x_p|$ for any $j<p$ satisfying  $|x_j-x_p|>\RR$.
	\item $R_p\geq \RR$
	\item When $\Theta^{in}_p=0$, there exists $j_p$ such that $|x_{j_p}-x_p|=R_p$
\end{enumerate} 
Similarly, we have
\begin{enumerate}
	\item $f(r_p)\leq f(x_j-x_p)$ for any $j<p$ satisfying  $|x_j-x_p|\leq \RR$,
	\item $r_p \leq |x_j-x_p|$ for any $j<p$ satisfying  $|x_j-x_p|\leq \RR$ and $f(x_j-x_p)=f(r_p)$.
	\item $r_p\leq \RR$
	\item When $\Theta^{out}_p=0$, there exists $i_p$ such that $|x_{i_p}-x_p|=r_p$
\end{enumerate} 
Then, we define a \textit{continuous} function $T$ on $\R$ as follows
\beq\label{defT}
T(|x|)=\left\{ \begin{array}{ll} 
1 &  2\RR\geq  |x| \\
(|x|^{-1}-b^{-1})(2\RR^{-1}-b^{-1})^{-1} & b\geq |x|\geq 2\RR \\ 
0 & { |x|\geq b}\,,
\end{array}\right. 
\eeq
At last we define $F_p(x_1,\cdots,x_N)$ on $[0,L]^{3N}$ as follows  ($1<p\leq N$),
\beqa
\label{deffp}
F_p\equiv \left\{ \begin{array}{ll} 
f(r_p) &  \Theta^{in}_p=1\\
f(R_p)&\Theta^{out}_p=1 \\ 
f(r_p)+T(R_p)\left[f(R_p)-f(r_p)\right]_ - & { otherwise}\,,
\end{array}\right. 
\eeqa
and $F_1=1$. Here $[\cdot]_-$ denotes the  negative part, i.e., $[x]_-=x$ when $x<0$ and $[x]_-=0$ when $x\geq 0$. We note that 
for any $x$
\beq
[x]_-\leq0
\eeq
\par Note: If $v\geq 0$, it is well known that $f$ is an increasing function, which implies the $F_p$ we defined is equal to the $F_p$ in \cite{dyson}.
\par One can prove that $F_p$ is a continuous function of $(x_1,\cdots,x_N)$ by checking that, for any $j\neq p>1$ and fixed $x_1,\cdots,x_{j-1},x_{j+1},\cdots,x_N$, $F_p$ is a continuous function of $x_j$. First we can see that it is trivial for $j>p$, since $F_p$ is independent of $x_j$ when $j>p$. For $j<p$, it only remains to check that $F_p$ is continuous when $x_j$ moves from $|x_j-x_p|=\RR$ to $|x_j-x_p|=\RR+0^+$.
One can see that when $|x_j-x_p|=\RR$, $f(R_p)\geq f(\RR)=f(x_j-x_p)\geq f(r_p)$, so $F_p=f(r_p)$, i.e. 
\beq
F_p=\min\bigg\{\min_{k:k\neq j, k< p}\{f(x_k-x_p):|x_k-x_p|\leq \RR\},f(x_j-x_p)\bigg\}
\eeq
On the other hand, when $|x_j-x_p|=\RR+0^+\leq 2\RR$, we can see that $R_p=|x_j-x_p|$, $T(R_p)=1$ and $f(R_p)=f(\RR)+0^+$. Hence, 
\beq
F_p=\min\bigg\{\min_{k:k\neq j, k< p}\{f(x_k-x_p):|x_k-x_p|\leq \RR\},f(x_j-x_p)\bigg\}
\eeq
Hence we arrive at the desired result that $F_p$ is continuous function.
\par We can also see that $F_p$ is non-negative and bounded as follows
 \beq
 M\equiv \|F_p\|_\infty=\|f\|_{\infty}\leq (1-a/b)^{-1}\|f_a\|_{\infty}=(1-a/b)^{-1}\|f_1\|_{\infty}\leq 2\|f_1\|_{\infty} .
 \eeq
Here we use the fact $f_a(x)=f_1(x/a)$. 
\par By the definition of $F_p$, one can see that $F_p=1$ when $\prod_{q<p}\bar\theta_b(x_p-x_q)=1$ and $F_p\leq 1$ when $\prod_{q<p}\bar\theta_\RR(x_p-x_q)=1$, so
\beq\label{fboundlu}
1-\sum_{q<p}\theta_b(x_p-x_q)\leq F_p\leq 1+\sum_{q<p}(M-1)\TR(x_p-x_q)
\eeq
\par We now construct the state functions $\Phi_k$ as follows ($1\leq k\leq N$)
 \[\Phi_k=\prod_{p=1}^kF_p\]
Note: all $\Phi$'s are functions on $[0,L]^{3N}$ and $\Phi_k$ is independent of $x_l$ for $l>k$. We will choose $\Psi=\Phi_N$ for (\ref{psiny}). 
\par As in \cite{LY1}, for proving the upper bound on the total energy $\langle\Phi_N|H_N|\Phi_N\rangle\|\Phi_N\|^{-2}_2$, we shall estimate the upper bounds on
\beq
\|\Phi_N\|^{-2}_2\int\sum_i|\nabla_i\Phi_N|^2\prod_{j=1}^Ndx_j\,\,\,\,\,\,{\rm and}\,\,\,\,\,\, \|\Phi_N\|^{-2}_2\int\sum_{i<j}v^a(x_i-x_j)|\Phi_N|^2 \prod_{k=1}^Ndx_k
\eeq
Since in our case $v^a$ has negative parts, our strategy is more complicated, i.e., we need to  estimate the upper bounds on 
\beq\label{ob1}
\|\Phi_N\|^{-2}_2\int\sum_i|\nabla_i\Phi_N|^2\prod_{j=1}^Ndx_j\,\,\,\,\,\,{\rm and}\,\,\,\,\,\, \|\Phi_N\|^{-2}_2\int\sum_{i<j}[v^a]_+(x_i-x_j)|\Phi_N|^2 \prod_{k=1}^Ndx_k\eeq
and the lower bound on 
\beq\label{ob2}
\|\Phi_N\|^{-2}_2\int\sum_{i<j}\bigg|[v^a]_-(x_i-x_j)\bigg|\cdot|\Phi_N|^2 \prod_{k=1}^Ndx_k
\eeq
In the remainder of this section we are going to prove the following three inequalities
\begin{itemize}
	\item $\|\Phi_N\|^{-2}_2\int\sum_i|\nabla_i\Phi_N|^2\prod_{j=1}^Ndx_j\leq (1+o(1))\frac{N^2}{\Lambda}\int_{\R^3}|\nabla f(x)|^2dx$
	\item $\|\Phi_N\|^{-2}_2\int\sum_{i<j}[v^a]_+(x_i-x_j)|\Phi_N|^2 \prod_{k=1}^Ndx_k\leq (1+o(1))\frac{N^2}{\Lambda}\int_{\R^3}\half [v]_+ |f(x)|^2dx$
	\item $\|\Phi_N\|^{-2}_2\int\sum_{i<j}\bigg|[v^a]_-(x_i-x_j)\bigg|\cdot|\Phi_N|^2 \prod_{k=1}^Ndx_k\geq (1-o(1))\frac{N^2}{\Lambda}\int_{\R^3}\half \left|[v]_-\right|\cdot|f(x)|^2dx.$
\end{itemize}
To prove these inequalities, we begin with proving the following three inequalities  (all $\Phi$'s are functions on $[0,L]^{3N}$): 
\begin{enumerate}
	\item 
	For any $m$-variable function $g_m(x_{i_1}\cdots x_{i_m})$, $m<k\leq N$, $i_j\neq i_k$ for $j\neq k$, we have
\beqa\label{str1}
&& \|\Phi_kF^{-1}_{i_1}\cdots F^{-1}_{i_m}g_m\|^2_2\leq (2M)^{2m}\Lambda^{-m}\|\Phi_{k-m}\|^2_2\|g_m\|^2_2
\eeqa
\item
For any two variable function $g_2(x_i,x_{i'})$ ($i<i'$), we have  
\beq\label{str2}
\|\Phi_NF^{-1}_{i'} g_2\|^2_2\leq \|g_2\|^2_2\Lambda^{-2}\|\Phi_N\|^2_2(1+\const Y)
\eeq
\item
Let $f_{i,i'}=f(x_i-x_{i'})$, for any two variable function $g_2(x_i,x_{i'})$ ($i<i'$),  we have  
\beqa\label{str3}
\|\Phi_Nf^{-1}_{i,i'} g_2\|^2_2\geq \|g_2\|^2_2\Lambda^{-2}\|\Phi_N\|^2_2(1-\const Y)
\eeqa
\end{enumerate}
Note: if $f_{i,i'}=0$ and $i< i'$, then $F_{i'}=0$, so $\Phi_Nf^{-1}_{i,i'}$ is definable.
\par  We will use (\ref{str1}) for controlling the error terms. The inequalities (\ref{str2}) and (\ref{str3}) will be used in estimating the terms (\ref{ob1}) and (\ref{ob2}), respectively.
\par We begin with deriving a lower bound on $\|\Phi_N\|^2_2$
. For $i\neq p$,  Let $F_{p,i}$ be the value that $F_p$ would take if changing the order of particles as follows, 
 \beq
F_{p,i}(x_1\cdots x_N)\equiv F_{n(p,i)}(x_1\cdots x_{i-1}, x_{i+1}\cdots x_N,x_i\,)
\eeq
Here $n(p,i)$ is defined as follows ($i\neq p$)
\begin{equation}
n(p,i)=  \left\{ \begin{array}{ll} 
p & i>p \\ 
p-1 & i<p\,,
\end{array}\right.
\end{equation}
Similarly, we can  define $F_{p,i,j}(x_1\cdots x_N)$ as 
\beq
F_{p,i,j}(x_1\cdots x_N)\equiv F_{m(p,i,j)}(x_1\cdots x_{i-1}, x_{i+1}\cdots x_{j-1}, x_{j+1}\cdots x_N,x_i,x_j\,)\,\,\, for\,\,\,i<j
\eeq
and $F_{p,i,j}=F_{p,j,i}$ for $j<i$. Here $m(p,i,j)$ is defined as the number of the elements of the set 
$
\{1,\cdots, p\}\setminus\{i,j\}.
$
\par Note: As we mentioned $F_p$ we defined is equal to the $F_p$ in \cite{LSY00} in the case when $v\geq0$. Furthermore, one can see that our definitions of $F_{p,i}$ and $F_{p,i,j}$ are equivalent to those definitions in \cite{LSY00} when $v\geq 0$.  
\par With the definitions of $F_p$ and $F_{p,i}$, we obtain that $F_{p,i}$ is independent of $x_i$ and $F_p$ is bounded from below as follows
\begin{equation}\label{flow}
F_p(x_1\cdots x_N)\geq  \left\{ \begin{array}{ll} 
F_{p,i} & i>p \\ 
F_{p,i}\bar\theta_b(x_p-x_i) & i<p\,,
\end{array}\right.
\end{equation}
and  
\[F_p\geq \prod_{i<p}\bar\theta_b(x_p-x_i).\]
Then  $\Phi^2_N$ is bounded from below, for any fixed $i$, by
\beqa
\bigg|F_{1}\cdot F_{2}\cdots F_{N}\bigg|^2&&\geq \bigg|F_{1,\,i}\cdots F_{i-1,\,i} F_{i+1,\,i}\cdots F_{N,\,i}\bigg|^2\times\prod_{j\neq i}\bar\theta_b(x_i-x_j)
\\\nonumber
&&\geq \bigg|F_{1,\,i}\cdots F_{i-1,\,i} F_{i+1,\,i}\cdots F_{N,\,i}\bigg|^2\times\bigg(1-\sum_{j\neq i}\theta_b(x_i-x_j)\bigg)
\eeqa
Integrating  both sides with $\int \prod_{j=1}^N dx_j$, we obtain that 
\beqa\label{psinnlow}
&&\|\Phi_N\|_2^2\geq \|\Phi_{N-1}\|_2^2\left(1-\frac{4\pi b^3}{3}N/\Lambda\right)=\|\Phi_{N-1}\|_2^2(1-Y)
\eeqa
Here we used the fact that 
\beq\label{n1ff}
\|\Phi_{N-1}\|^2_2=\int \bigg|F_{1,\,i}\cdots F_{i-1,\,i} F_{i+1,\,i}\cdots F_{N,\,i}\bigg|^2\prod_{j}dx_j.
\eeq 
Similarly, one can also prove that for $k\leq N$, 
\beqa\label{n1ff2}
\|\Phi_k\|_2^2\geq \|\Phi_{k-1}\|_2^2(1-Y)
\eeqa
\par Next we are going to prove (\ref{str1}) in the case $m=1,k=N$, i.e., 
\beqa\label{inonegu}
\| \Phi_NF_i^{-1}g_1\|^2_2\leq 4M^{2}\Lambda^{-1}\| \Phi_{N-1}\|^2_2\,\| g_1\|^2_2
\eeqa
One can check that
  $F_p>F_{p,i}$ only when the following conditions are satisfied:
\begin{enumerate}
\item $i<p$
	\item $|x_i-x_p|\leq  \RR$,
	\item for any other $j<p$, $|x_j-x_p|$ is greater than $\RR$,
	\item $T(R_p)< 1$, i.e. for any other $j<p$, $|x_j-x_p|>2\RR$,
\end{enumerate} 
i.e.,
\beq\label{defgpi} 
F_p>F_{p,i}\Rightarrow
G_{p,i}\equiv\TR(x_p-x_i)\prod_{j< p, j\neq i}\bar\theta_{2\RR}(x_j-x_p)=1
\eeq 
On the other hand, 
using the fact that $f(\RR)>\half$, one obtains that if $F_p>F_{p,i}$, 
\[F_{p,i}\prod_{j< p, j\neq i}\bar\theta_{2\RR}(x_j-x_p)>f(\RR)\prod_{j< p, j\neq i}\bar\theta_{2\RR}(x_j-x_p)\geq\frac12\prod_{j< p, j\neq i}\bar\theta_{2\RR}(x_j-x_p)\] 
Hence when $G_{p,i}=1$, 
we have  $2MF_{p,i}G_{p,i}\geq M\geq F_p$, i.e., 
\beq\label{prfp1}
F_p\leq  \left\{ \begin{array}{ll} 
F_{p,i} & i>p \\ 
F_{p,i}\bigg(1+(2M-1)G_{p,i}\bigg) & i<p\,.
\end{array}\right.
\end{equation}
By the definition of $G$'s, one can see that if $p,q>i$ and $p\neq q$, 
\beq\label{progpq}
G_{p,i}G_{q,i}=0.
\eeq
Hence, we have that
\beqa\label{ffff1}
\prod_{p>i}\bigg(1+(2M-1)G_{p,i}\bigg)\leq  2M
\eeqa
Combining (\ref{prfp1}) and (\ref{ffff1}), we have the upper bound on $|\Phi_N F_{i}^{-1}|$ as follows,
\beqa
\bigg|F_{1}\cdot F_{2}\cdots F_{N}F_{i}^{-1}\bigg|^2\leq 4M^{2} \bigg|F_{1,\,i}\cdots F_{i-1,\,i} F_{i+1,\,i}\cdots F_{N,\,i}\bigg|^2,
\eeqa
which implies the desired result (\ref{inonegu}) with (\ref{n1ff}).
Furthermore, for any $m$-variable function $g_m(x_{i_1}\cdots x_{i_m})$
\beqa\label{spe3}
&& \|\Phi_NF^{-1}_{i_1}\cdots F^{-1}_{i_m}g_m\|^2_2\leq (2M)^{2m}\Lambda^{-m}\|\Phi_{N-m}\|^2_2\|g_m\|^2_2
\eeqa
With the inequality (\ref{psinnlow}) and the fact $F_i\leq M$ for any $i\leq N$, we get
\beqa\label{1314}
 \|\Phi_Ng_m\|^2_2\leq &&(2M^2)^{2m}\|\Phi_{N-m}\|^2_2\|g_m\|^2_2\\\nonumber\leq&& (1-Y)^{-m}(2M^2)^{2m}{\Lambda}^{-m}\|\Phi_{N}\|^2_2\|g_m\|^2_2
\eeqa
Similarly, we can generalize this result to $m<k\leq N$
\beqa\label{13142}
 \|\Phi_kg_m\|^2_2\leq &&(2M^2)^{2m}\|\Phi_{k-m}\|^2_2\|g_m\|^2_2\\\nonumber\leq&& (1-Y)^{-m}(2M^2)^{2m}{\Lambda}^{-m}\|\Phi_{k}\|^2_2\|g_m\|^2_2
\eeqa
Now we shall prove  the upper bound on $\|\Phi_N\|^2_2$ with (\ref{13142}). Choosing $p=N$, with the bounds of $F_p$ in (\ref{fboundlu}), we get that
\beqa
\Phi_N^2\leq&& F^2_{1}\cdot F^2_{2}\cdots F^2_{N-1}\left(1+\sum_{j<N}M^2\TR(x_j-x_N)\right)\\\nonumber
=&&\Phi_{N-1}^2+\Phi_{N-1}^2\sum_{j<N}M^2\TR(x_j-x_N)
\eeqa
Hence, using the inequalities (\ref{13142})($m=1$) and (\ref{n1ff2}), we obtain that 
\beqa\label{psinnupper}
\|\Phi_N\|^2_2&&\leq \|\Phi_{N-1}\|^2_2 +(1-Y)^{-1}(2M^2)^2Y\|\Phi_{N-2}\|^2_2\\\nonumber
&&\leq \|\Phi_{N-1}\|^2_2(1+\const Y)
\eeqa
Putting (\ref{psinnupper}) and (\ref{psinnlow}) together, we obtain the relation between $\|\Phi_N\|$ and $\|\Phi_{N-1}\|$ 
\beq\label{psinneq}
\|\Phi_N\|^2_2=\|\Phi_{N-1}\|^2_2(1+O(Y))
\eeq
Similarly, for $k\leq N$
\beq\label{psinneq2}
\|\Phi_k\|^2_2=\|\Phi_{k-1}\|^2_2(1+O(Y))
\eeq
\par Next, we shall prove (\ref{str2}), i.e., 
\beqa
\|\Phi_NF^{-1}_{i'} g_2\|^2_2\leq \|g_2\|^2_2\Lambda^{-2}\|\Phi_N\|^2_2(1+\const Y).
\eeqa
Using the inequalities $F_i\leq \left(1+\sum_{l<i}(M-1)\TR(x_l-x_i)\right)$ and (\ref{prfp1}), with the property of the $G$'s in (\ref{progpq}), we get
\beqa\label{ffff2}
\prod_{k\neq i'}F_{k}
&&\leq \left(1+\sum_{l<i}(M-1)\TR(x_l-x_i)\right)\prod_{k\neq i',i}F_{k\,,i}\cdot\bigg(1+\sum_{j>i} 2MG_{j,i}
\bigg)
\eeqa
Similarly, replacing $F_{p,i}$'s with $F_{p,i,i'}$'s and using the fact that $G_{k,l}\leq \TR(x_k-x_l)$, we get 
\beqa\label{wget}
\Phi_NF^{-1}_{i'}\leq &&\prod_{k\neq i',i}F_{k,i,i'}\left(1+\sum_{l<i}M\TR(x_l-x_i)\right)\\\nonumber
&&\times \bigg(1+\sum_{j>i} 2M\TR(x_j-x_i)\bigg)\times \bigg(1+\sum_{j'>i'} 2M\TR(x_j'-x_i')\bigg)
\eeqa 
Expanding (\ref{wget}), multiplying $g_2(x_i,x_{i'})$ to each side and integrating them with $\prod_{k=1}^N dx_k$, with the result of (\ref{13142},  \ref{psinneq2}), we obtain that 
\beq\label{speleq}
\|\Phi_NF^{-1}_{i'} g_2\|^2_2\leq \|g_2\|^2_2\Lambda^{-2}\|\Phi_N\|^2_2(1+\const Y)
\eeq
\par So far we proved some upper bounds of the expectation value of $\Phi_N$. Next we are going to prove the following lower bound on  $\|\Phi_Nf^{-1}_{i,i'} g_2\|^2_2$:
\beqa\label{spegeq}
\|\Phi_Nf^{-1}_{i,i'} g_2\|^2_2\geq \|g_2\|^2_2\Lambda^{-2}\|\Phi_N\|^2_2(1-\const Y)
\eeqa
Here we denote $f_{i,i'}= f(x_i-x_{i'})$ ($i<i'$).  
First, by the definition of $F_{i'}$, one can see that $F_{i'}=f(x_i-x_{i'})$ when $\prod_{k<i',k\neq i}\bar\theta_b(x_k-x_i')=1$, i.e.,  
\beq F^2_{i'}\geq f(x_i-x_{i'})^2\left(1-\sum_{k<i',k\neq i}\theta_b(x_k-x_{i'})\right)\eeq
Using this inequality and (\ref{spe3}) with $m=3$, $i_1=i$, $i_2=i'$ and $i_3=k$, we obtain that
\beqa
\left\|\Phi_Nf^{-1}_{i,i'} g_2\right\|^2_2\geq \left\|\Phi_NF^{-1}_{i'} g_2\right\|^2_2
-\const Y\left\| g_2\right\|^2_2 \Lambda^{-2}\left\|\Phi_N\right\|^2_2\eeqa
Then with the lower bound on  $F_i^2$ in (\ref{fboundlu}), i.e., $F^2_{i}\geq 1-\sum_{k<i}\theta_b(x_k-x_{i})$, 
we obtain that 
\beqa
\|\Phi_Nf^{-1}_{i,i'} g_2\|^2_2\geq \|\Phi_NF^{-1}_{i}F^{-1}_{i'} g_2\|^2_2
-\const Y\| g_2\|^2_2 \Lambda^{-2}\|\Phi_N\|^2_2
\eeqa 
Again, using the bound on  $F_p$ in (\ref{flow}), we see that
\beqa\nonumber
\Phi_NF^{-1}_{i}F^{-1}_{i'}\geq&& \prod_{k\neq i,i'}F_{k,i,i'}\left(1-\sum_{l<i}\theta_b(x_l-x_i)\right)\times \left(1-\sum_{l'<{i'},\,\,l'\neq i}\theta_b(x_{l'}-x_{i'})\right)
\eeqa
Then using (\ref{13142}) and (\ref{psinneq2}),  we arrive at  the desired result (\ref{spegeq}). 
\par  
So far, we have proved the inequalities we need for calculating the value of $\langle \Phi_N|\sum_{i,j}v^a(x_i-x_j)|\Phi_N\rangle$. Then we need to calculate $\nabla_i \Phi_N$. We denote $i_p$ as the particle satisfying ${i_p}<p$ and $|x_{i_p}-x_p|=r_p$ and $n^r_p$ as the unit vector in the direction of $x_p-x_{i_p}$. Similarly, denote $j_p$ as the particle satisfying ${j_p}<p$ and $|x_{j_p}-x_p|=R_p$ and $n^R_p$ as the unit vector in the direction of $x_p-x_{j_p}$. 
We remark that such $i_p$ or $j_p$ may not exist in some cases, but we do define them as $0$. We denote $\nabla_0F_p=0$. Recall the definition of $F_p$ in (\ref{deffp}). We have 
\beqa\label{fffp}
-\nabla_p F_p&&=\nabla_{i_p} F_p+\nabla_{j_p} F_p\\\nonumber
\nabla_{i_p} F_p&&= -n^r_p f\,'(r_p)\bigg(\Theta^{in}_p +\Theta^-_p(1-T(R_p))+\Theta^+_p\bigg)\\\nonumber
\nabla_{j_p} F_p&&=-n^R_p\bigg( \Theta^{out}_p f\,'(R_p)+\Theta^-_pT(R_p)f'(R_p)+\Theta^-_pT\,'(R_p)\big(f(R_p)-f(r_p)\big)\bigg)
\eeqa
Here $\Theta^+_p$ is the function of $(x_1\cdots x_N)$ which is defined as
\beq
\Theta^+_p\equiv \big[1-\Theta^{in}-\Theta^{out}\big] \cdot h\big[f(R_p)- f(r_p)\big]
\eeq 
and $\Theta^-_p$ is defined as
\beq
\Theta^-_p\equiv \big[1-\Theta^{in}-\Theta^{out}\big] \cdot  h\big[f(r_p)- f(R_p)\big]
\eeq
Here $h$ is the Heaviside step function. By the definition of $\Phi_N$, we obtain that
\beqa\nonumber
&&\frac{|\nabla_p\Phi_N|^2}{|\Phi_N|^2}=\left|-F_p^{-1}\nabla_{i_p}F_p-F_p^{-1}\nabla_{j_p}F_p+\sum_{q,i_q=p}F_q^{-1}\nabla_{p}F_q+\sum_{q,j_q=p}F_q^{-1}\nabla_{p}F_q\right|^2
\eeqa
Then with (\ref{fffp}), we have that
\beqa\nonumber
\sum_p|\nabla_p\Phi_N|^2\leq &&2|\Phi_N|^2\sum_{p}F_p^{-2}\bigg(|f\,'(r_p)|^2\big(\Theta^{in}_p +\Theta^-_p|1-T(R_p)|^2
+\Theta^+_p\big)\\\nonumber&&+ |T(R_p)f\,'(R_p)|^2\Theta^-_p+|T\,'(R_p)|^2|f(R_p)-f(r_p)|^2\Theta^-_p\\\nonumber
&&+|T(R_p)|\cdot|1-T(R_p)|\cdot|f\,'(r_p)|\cdot|f\,'(R_p)|\Theta^-_p+|f\,'(R_p)|^2\Theta^{out}_p\bigg)\\\nonumber
&&+2|\Phi_N|^2\sum_{k<p<q}F_p^{-1}F_q^{-1}\bigg(|\nabla_kF_p|\cdot|\nabla_pF_q|+|\nabla_kF_p|\cdot|\nabla_kF_q|\bigg)
\eeqa
Because $0\leq T\leq 1$ and  $i_p\neq j_p$, one can easily prove that for any fixed $p$,
\beqa\nonumber
&&\bigg(|f\,'(r_p)|^2\big(\Theta^{in}_p +\Theta^-_p|1-T(R_p)|^2
+\Theta^+_p\big)+|f\,'(R_p)|^2\Theta^{out}_p\\\nonumber+&& |T(R_p)f\,'(R_p)|^2\Theta^-_p+|T(R_p)|\cdot|1-T(R_p)|\cdot|f\,'(r_p)|\cdot|f\,'(R_p)|\Theta^-_p\bigg)\\\nonumber
\leq&& \sum_{k:k< p} f\,'(|x_p-x_k|)^2,
\eeqa
and 
\beq
|T\,'(R_p)|^2|f(R_p)-f(r_p)|^2\Theta^-_p\leq M^2\sum_{k:k< p}\left( T\,'(|x_p-x_k|)^2\sum_{j:j\neq k,\, p}\TR(x_j-x_p) \right)
\eeq
Hence,  we obtain that 
\beqa\nonumber
&&\langle \Phi_N|H_N|\Phi_N\rangle\leq 2\sum_{i<j}{\int |\Phi_N|^2 F_j^{-2}\bigg(\half f(x_i-x_j)^2[v(x_i-x_j)]_++f\,'(x_i-x_j)^2\bigg)}\\\nonumber
&&-2\sum_{i<j}{\int |\Phi_N|^2 f^{-2}(x_i-x_j)\bigg(\half f^{2}(x_i-x_j) \bigg|[v(x_i-x_j)]_-\bigg|\bigg)}\\\label{inequhn}
&&+2\sum_{i<j}\int |\Phi_N|^2M^2\sum_{k< p}\left( T\,'(|x_p-x_k|)^2\sum_{j:j\neq k,\, p}\TR(x_j-x_p) \right)\\\nonumber
&&+2\sum_{k<p<q}\int |\Phi_N|^2F_p^{-1}F_q^{-1}\bigg(|\nabla_kF_p|\cdot|\nabla_pF_q|+|\nabla_kF_p|\cdot|\nabla_kF_q|\bigg)
\eeqa
Here $[\cdot]_+$ and $[\cdot]_-$ denote the positive and negative part, respectively and we used the fact that $F_j\leq f(x_i-x_j)$ when $i<j$ and $|x_i-x_j|\leq \RR$, which implies that  
\beq
[v(x_i-x_j)]_+\leq F_j^{-2}f(x_i-x_j)^2[v(x_i-x_j)]_+
\eeq
 With the results in (\ref{speleq}) and (\ref{spegeq}), we can obtain the upper bound on  the main part of $\langle \Phi_N|H_N|\Phi_N\rangle$, i.e.,
\beqa\nonumber
&&2\sum_{i<j}\int |\Phi_N|^2\times \\\nonumber
&&\frac12\bigg(F_j^{-2}\left[ f^2(x_i-x_j)[v(x_i-x_j)]_++f\,'(x_i-x_j)^2\right]- \frac{f^2(x_i-x_j)}{f^{2}(x_i-x_j)} \bigg|[v(x_i-x_j)]_-\bigg|\bigg)\\\label{mapaof}
&&\leq 4\pi a N^2/\Lambda(1+\const Y)\|\Phi_N\|^2_2
\eeqa
With the definition of $T$ in (\ref{defT}) and (\ref{1314}), we obtain that the third line of (\ref{inequhn}) is bounded as $\const aN^2 Y\|\Phi_N\|^2_2/\Lambda$. For the other terms, we have
\beq
|\nabla_{i_p}F_p|\leq |f\,'(|x_{i_p}-x_p|)|,\,\,\,\,|\nabla_{j_p}F_p|\leq |f\,'(|x_{j_p}-x_p|)|+MT\,'(|x_{j_p}-x_p|)
\eeq
Hence, with the inequality (\ref{spe3}), we can prove that the last line in (\ref{inequhn}) are bounded as
\beq
\const N^3\Lambda^{-2}(K+L)^2\|\Phi_N\|^2_2
\eeq
Here $K$ and  $L$ are defined as follows 
\beqa
K\equiv \int_{ \R^3} |f\,'\left(|x-y|\right)|dy\,\,\,\,\,\,\,L\equiv \int_{\R^3} T\,'\left(|x-y|\right)dy.
\eeqa
Note that  $K$ and $L$ are independent of $x$. 
By the definitions of $f$ in (\ref{deff}) and $T$ in (\ref{defT}), we get that 
\beq
K=O(ab),\,\,\,\ L=O(\RR b)=O(a b)
\eeq
 Hence we obtain that the last line in (\ref{inequhn}) are bounded by $\const a N^2Y^2$. Combining this result with    (\ref{mapaof}), we get the following result,
\beq
\frac{\langle \Phi_N|H_N|\Phi_N\rangle}{\|\Phi_N\|^2_2}\leq  4\pi a N^2/\Lambda(1+\const Y)
\eeq
At last, by choosing $\Psi=\Phi_N$, we arrive at the desired result (\ref{psiny}), which implies Theorem 1.  
\end{proof}

\subsection{Proof of Theorem Two}
\begin{proof}
\par Following the ideas in \cite{LY1}, we need to replace the hard potential by a \textit{soft} potential at the expense of local kinetic energy. This method has been used in many papers on dilute bose or fermi gases \cite{LY1, LY2, LSY00, LS1, LSS}. But in this method the kinetic energy of particle $i$ only can be used for the hard-soft potential replacement between the particle $i$ and \textit{one }other $j$ (the nearest particle  \cite{LY1}). In our case that $v^a$ is partly negative,  we can not  ignore the potential between $i$ and other $k$'s for the lower bound on  the energy. To solve this problem, we begin with separating the whole Hamiltonian into two parts, (1) The Hamiltonian of the energy when two particles are close to each other and they are far away from the others. (2) The Hamiltonian of the remaining energy. In the remainder of this section, we prove that the first part is greater than $4\pi aN^2\Lambda^{-1}(1-O(a^3\rho)^{1/17})$ and the second part is non-negative.
\par Another important property Lieb and Yngvason used in \cite{LY1} is the superadditivity of the ground energy $E(n,\ell)$ of $n$ particles in $[0,\ell]^3$ with Neumann boundary condition, i.e.,
\beq
E(n+n',\ell)\geq E(n,\ell)+E(n',\ell)
\eeq  
This property is trivial in the case $v^a\geq0$. In our proof, we are not going to prove any similar property, actually we only need the property (\ref{enell2}) that for fixed $\ell$, when $n$ is larger than $4\rho\ell^3$, the energy/particle is greater than $8\pi a \rho$, as in (2.62) of \cite{ERJJ2}, i.e., 
\beq
E(n,\ell)/n\geq 8\pi a\rho(1-\const (a^3\rho)^{1/17})
\eeq
which will be proved in Lemma 1.
\par Choosing 
\beq 
\label{defR}R=a (a^3\rho)^{-5/17}\geq2R_0a,\eeq 
  we define $F_{i,j}$ for $i\neq j$ as follows:
\beq\label{defFij}
F_{i,j}=\theta_R(x_i-x_j)\prod_{k\neq i,j}\bar\theta_{2R}(x_i-x_k)
\eeq
Here $\theta_R$ is the characteristic function of the open set $|x|<R$, and $\bar\theta_R=1-\theta_R$. 
We note that $F_{i,j}\neq F_{j,i}$ and $F_{i,j}$ is equal to $1$ only when $x_j$ is close to $x_i$, but the other $x_k$'s are not.  It is easy to check that $\sum_{i:i\neq j}F_{i,j}\leq 1$, so
\beq
-\nabla_j\sum_{i:i\neq j}F_{i,j}\nabla_j\leq -\Delta_j
\eeq
for any fixed $x_1,\cdots x_{j-1},x_{j+1},\cdots, x_N$. 
\par Then we denote  $v_+^a$ and $v_-^a$ as scaled potentials as follows,
\beq\label{defv+a}
v_+^a(r)=a^{-2}v_+(r/a),\,\,\,v_-^a(r)=a^{-2}v_-(r/a)
\eeq
Choosing 
\beq\label{defY}
Y=(a^3\rho)^{1/17}
\eeq and $\eps$ satisfying
\beq
\label{defeps} 3\cdot \bigg(\min\big\{1,SL[v_+]\big\}\bigg)^{-1}\cdot Y= \eps<\frac t{2(1+t)},
\eeq
with the definition 
\beq\label{defveps}
v_\eps^a\equiv v^a-\eps v^a_+,
\eeq
 we separate the  Hamiltonian $H_N$ as follows
\beqa \label{seHN}
&&H_N=\\\nonumber
&&(1-\eps)\sum_{j}-\nabla_j\sum_{i}F_{i,j}\nabla_j+\sum_{i\neq j}F_{ij}
\frac{v_\eps^a}{2}(x_i-x_j)-\eps\sum_{j}\Delta_j+\eps\sum_{i\neq j} \frac{v^a_+}{2}(x_i-x_j)\\\nonumber
&&+(1-\eps)\sum_{j}-\nabla_j(1-\sum_{i}F_{i,j})\nabla_j+\sum_{i\neq j}(1-F_{ij})\frac{v_\eps^a}{2}(x_i-x_j)
\eeqa
First, we claim the following Lemma 1, which will be proved in next section. 
\begin{lem} Define $Y$, $F_{i,j}$, $\eps$, $v^a_\eps$ and $R$ as in (\ref{defY}), (\ref{defFij}), (\ref{defeps}), (\ref{defveps})  and (\ref{defR}) respectively.  There exists $C$ depending only on  $v$ such that
\beqa\label{lemone}
&&H''\equiv\\\nonumber
 &&(1-\eps)\sum_{j}-\nabla_j\sum_{i}F_{i,j}\nabla_j+\sum_{i\neq j}F_{ij}
\frac{v_\eps^a}{2}(x_i-x_j)-\eps\sum_{j}\Delta_j+\eps\sum_{i\neq j} \frac{v^a_+}{2}(x_i-x_j)
\\\nonumber
&&\geq 4\pi aN^2/\Lambda(1-CY)
\eeqa 
\end{lem}
Hence, to obtain Theorem 2, it only remains to prove that  the last line of (\ref{seHN}), as an operator,  is bounded from below by zero, i.e., 
\[(1-\eps)\sum_{j}-\nabla_j(1-\sum_{i}F_{i,j})\nabla_j+\sum_{i\neq j}(1-F_{ij})\half v^a_\eps(x_i-x_j)\geq 0\]
By the assumptions $\eps<\,t\,(2+2t)^{-1}$, we have  
\[v_\eps^a\geq \frac{2+t}{2+2\,t}v^a_++ v^a_-.\] 
Hence, it remains to prove that 
\beq\nonumber
0\leq H_N'\equiv \frac{2+t}{2+2\,t}\sum_{j}-\nabla_j(1-\sum_{i}F_{i,j})\nabla_j+\frac12\sum_{i\neq j}(1-F_{ij})\left(\frac{2+t}{2+2\,t}v^a_++ v^a_-\right)(x_i-x_j)
\eeq
Because $\lim_{N\to\infty} E(N,\Lambda)/N$ exists, for proving Theorem 2, we can assume that $N$ is even, i.e., $N=2N_1$. Consider any partition
$P = (\pi_1,\pi_2)$ of ${1, ...,N}$ into two disjoint sets with $N_1$ integers in $\pi_1$ and $\pi_2$ respectively. For each $P$, we define that 
\beqa\nonumber
H_P= H_{(\pi_1,\pi_2)}\equiv&&\frac{2+t}{1+t}\sum_{j\in\pi_1}-\nabla_j(1-\sum_{i\neq j}F_{i,j})\nabla_j+\sum_{i,j\in\pi_1}(1-F_{i,j})\half v_{1,1}^a(x_i-x_j)\\\nonumber
+&&\sum_{i\in\pi_2, j\in\pi_1}(1-F_{i,j})\half v^a_{2,1}(x_i-x_j)+ \sum_{i,j\in\pi_2}(1-F_{i,j})\half v^a_{2,2}(x_i-x_j)
\eeqa
Here we denote  $v^a_{\al,\beta}$ as the interaction potential between particles in $\pi_\al$ and $\pi_\beta$, which are chosen as 
 \beq\label{defv1122}
v_{1,1}^a=v^a_{2,2}=\frac t{1+t}v^a_+\geq 0,\,\,\,\,v^a_{2,1}=\frac4{1+t}v^a_++4v^a_-,
\eeq 
so 
\[\frac14\big(v^a_{1,1,}+v^a_{2,1}+v^a_{2,2}\big)=\frac{2+t}{2+2\,t}v^a_++ v^a_-\leq v^a_\eps.\]
It is easily to check that 
\beq
H'_N=\sum_PH_P/\sum_P 1
\eeq
Hence, to obtain $H'_N\geq 0$, it remains to prove that for  $\forall P$, $H_P\geq 0$. Because there is no kinetic energy of particles in $\pi_2$, we can fix the configuration of $x_{i}$'s with $i\in\pi_2$. 
Since permutation of the labels in $\pi_1$ and $\pi_2$ is irrelevant, we assume that $\pi_1=\{1,\cdots,N_1\}$, $\pi_2=\{N_1+1,\cdots,N\}$. 
\par As we can see  $v^a_{2,1}$ is the only partly negative component in $H_P$. For fixed $\pi_2$ particles, we can write $v^a_{2,1}(x_j-x_i)$ as 
\beq\label{sev21}
v^a_{2,1}(x_j-x_i)=v^a_{2,1}(x_j-x_i)(1-\chi_A(x_i))+v^a_{2,1}(x_j-x_i)\chi_A(x_i)
\eeq
 Here $\chi_A$ is the characteristic function of $A$, which is a subset of $[0,L]^3$ (\ref{defA}). We shall show $A$ is the area where the density of $\pi_2$ particles is less than some fixed number.  To obtain $H_P\geq 0$, our strategy is to prove that 
\begin{enumerate}
	\item The total energy of the interaction potential $v_{1,1}^a$ and $v^a_{2,2}$ cancels out the negative part of $v^a_{2,1}(1-\chi_A)$.
	\item The total kinetic energy and the positive part of $v^a_{2,1}$ cancels out the negative part of  $v^a_{2,1}\chi_A$.
\end{enumerate}
\par 
To make the strategy more clear, we shall define $A$ where the density of $\pi_2$ particles is less than some fixed number. First we  
divide the cubic box $[0,L]^3$ into small cubes  $B_n$ ($n\in\N$) of side length $\ell$, with
  \[\ell= \half r_1a.\]
 Then, with fixed $x_{i}$'s, $i\in\pi_2$, for any $x\in [0,L]^3$,  we define the $G(x)$ as the set of $i$'s which satisfy $i\in\pi_2$ and $|x_i-x|\leq R_0a$, i.e., 
\beqa\label{defGx}
G(x)\equiv\{i\in \pi_2: |x_i-x|\leq R_0a\}
\eeqa
We denote $|G(x)|$ as the number of the elements of $G(x)$. 
\par
 We denote $d(x, B_n)$ as the distance between the cube $B_n\subset \R^3$ and $x\in\R^3$. Since $|G(y)|$ is uniformly bounded ($|G(y)|\leq N_1$), 
there must exist a point $X(B_n)\in\R^3$ satisfying $d(X(B_n), B_n)\leq 2R_0a$ and 
\beqa\label{defGb}
|G(X(B_n))|=\max\{|G(y)|: d(y, B_n)\leq 2R_0a  \}
\eeqa
We define $G(B_n)\equiv G(X(B_n))$. We are going to  prove that there exists $n_1\in \N$ depending on $R_0/r_1$ such that 
\begin{enumerate}
	\item The total energy of the interaction potential $v_{1,1}^a$ and $v^a_{2,2}$ cancels out the  negative parts of $v^a_{2,1}(x_j,x_i)$'s when $x_i$ is in a cube $B_n$ such that $|G(B_n)|> n_1$.
	\item The total kinetic energy and the positive part of $v^a_{2,1}$ cancel out the negative part of the remaining $v^a_{2,1}$'s.
\end{enumerate}

\par 
First, we derive the lower bound on the total energy of $v^a_{2,2}$, i.e. (\ref{48}, \ref{ij2}).  With the definition of $G(B_n)=G(X(B_n))$, we know that the set $\{x_k: k\in G(B_n)\}$ can be covered by a sphere of radius $R_0a$. So the number of the cubes which one need to cover this set is less than $\const (R_0/r_1)^3$. We denote these cubes as $B_{n_1}\cdots B_{n_m}$ $(m\leq \const (R_0/r_1)^3)$ and  assume the number of $i$'s satisfying $i\in G(B_n)$ and $x_i\in B_{n_k}$ is $a_{n_k}$. Because the side length of $B_{n_k}$ is equal to  $r_1a/2$, the distance between the two particles in the same cube is no more  than $\frac{\sqrt 3}2r_1a<r_1a $. Hence we have 
\beqa
\sum_{i,j\in G(B_n)}\theta_{r_1a}(x_i-x_j)\geq&& \sum_{k=1}^m \left[(a_{n_k})^2-(a_{n_k})\right]\\\nonumber
\geq &&\frac{(\sum_{k=1}^m a_{n_k})^2}{m}-(\sum_{k=1}^m a_{n_k})\\\nonumber
\geq &&\const (R_0/r_1)^{-3}|G(B_n)|^2-|G(B_n)|
\eeqa
Hence, we obtain that there exist $n_1\geq 3$ and  $n_1,n_2=\const (R_0/r_1)^3$ such that when $|G(B_n)|\geq n_1$, 
\beq
\sum_{i,j\in G(B_n)}\theta_{r_1a}(x_i-x_j)\geq \frac{1}{n_2}|G(B_n)|^2,
\eeq
which implies
\beq
\sum_{i,j\in G(B_n)}v^a_{2,2}(x_i-x_j)\geq \frac{t\lambda_+a^{-2}}{(1+t)n_2}|G(B_n)|^2
\eeq
Here, we used (\ref{defv1122}) and (\ref{defv+a}), i.e., 
\beq
v^a_{2,2}(r)=\frac{t}{(1+t)}v_+^a(r)=\frac{t}{(1+t)}a^{-2}v_+(r/a)
\eeq
Again, with the fact that the set $\{x_k: k\in G(B_n)\}$ can be covered with a sphere of diameter $2R_0a\leq R$, one can see that if $i\in G(B_n)$ and $|G(B_n)|\geq 3$, we have $F_{i,j}=0$ for any $j\neq i$. Hence we obtain that, for any fixed $B_n$ satisfying $|G(B_n)|\geq n_1$,
\beq\label{48}
\sum_{i,j\in G(B_n)}(1-F_{i,j})v^a_{2,2}(x_i-x_j)=\sum_{i,j\in G(B_n)}v^a_{2,2}(x_i-x_j)\geq \frac{t\lambda_+a^{-2}}{(1+t)n_2}|G(B_n)|^2
\eeq 
\par Then, we are going to sum up all the cubes satisfying $|G(B_n)|\geq n_1$. It is easy to see that
\beq
 d(x_i,B_n)\leq 3R_0a,\,\,\,\,{\rm{for}} \,\,\,\, i\in G(B_n),
\eeq
which implies that for any fixed $i\in {\pi_2}$, the number of cubes $B_n$'s satisfying  $i\in G(B_n)$ is  less than some constant $n_3$, which is less than $\const(R_0/r_1)^3$. Hence, summing up all the blocks satisfying $|G(B_n)|\geq n_1$, with the inequality (\ref{48}), we get that 
\beqa\nonumber
\sum_{i,j\in \pi_2}(1-F_{i,j})v^a_{2,2}(x_i-x_j)&&\geq \sum_{n:|G(B_n)|\geq n_1}\sum_{i,j\in G(B_n)}(1-F_{i,j})v^a_{2,2}(x_i-x_j)\\\label{ij2}
&&\geq\sum_{n:|G(B_n)|\geq n_1}\frac{t\lambda_+a^{-2}}{(1+t)n_2 n_3}|G(B_n)|^2
\eeqa
Second, we derive the lower bound on  the interaction potential between particles in $\pi_1$. Because the distance between any two points in the same cube is less than $r_1a$, we have $v_{1,1}^a(x_i-x_j)\geq a^{-2}\lambda_+ t(1+t)^{-1}$ when $i,j\in\pi_1$ and $x_i, x_j\in B_n$, i.e.,  
\beq\label{50.1}
\sum_{i,j\in\Pi_1(B_n)}v_{1,1}^a(x_i-x_j)\geq \frac{a^{-2}t\lambda_+}{1+t}\bigg(|\Pi_1(B_n)|^2-|\Pi_1(B_n)|\bigg) 
\eeq
Here $\Pi_1(B_n)$ is defined as the set of $i$'s such that  $i\in\pi_1$ and $x_i\in B_n$ and $|\Pi_1(B_n)|$ is the number of the elements of $\Pi_1(B_n)$. Furthermore, if $x_i\in B_n$ and $|G(B_n)|\geq 1$, there must be a $k\in\pi_2$ satisfying $|x_i-x_k|\leq 4R_0a\leq 2R$, hence $F_{i,j}=0$ for any other $j\in\pi_1$. Using this result, for any $B_n$ satisfying $|G(B_n)|\geq 1$, we have that 
\beq\label{50}
\sum_{i,j\in\Pi_1(B_n)}(1-F_{i,j})v_{1,1}^a(x_i-x_j)\geq \frac{t\lambda_+a^{-2}}{1+t}\bigg(|\Pi_1(B_n)|^2-|\Pi_1(B_n)|\bigg) 
\eeq
At last, we derive the lower bound on $v^a_{2,1}$. By the definitions of $|G(B_n)|$ and $v^a_{2,1}$, we have that  $\forall x\in B_n$,
 \[\sum_{i\in\pi_2}[v^a_{2,1}]_-(x- x_i)\geq -4\lambda_-a^{-2}|G(B_n)|.\]
Here we denote $[v^a_{2,1}]_-$ as the negative part of $v^a_{2,1}$ which is equal to $ 4 [v^a]_-$.  With the facts  $0\geq 4 [v^a]_-\geq-4 \lambda_-a^{-2}$ and $0\leq F_{i,j}\leq 1$, we have the following inequality
\beq\label{51}
\sum_{j\in\Pi_1(B_n),\,\, i\in\pi_2}(1-F_{i,j})[v^a_{2,1}]_-(x_i-x_j)\geq -4\lambda_-a^{-2}\cdot|\Pi_1(B_n)|\cdot|G(B_n)|
\eeq
One can check that if $|G(B_n)|\geq n_1$ and 
\beq\label{lambda+-}
\lambda_+\geq (1+t^{-1})\lambda_-\cdot\max\{2\sqrt{n_2n_3},\frac{n_2n_3}{4n_1}\} \sim \const(1+t^{-1})\lambda_-(R_0/r_1)^3,
\eeq
the sum of the right sides of (\ref{50}) and (\ref{51}) is bounded from below as follows,
\beqa
 &&-4\lambda_-\cdot|\Pi_1(B_n)|\cdot|G(B_n)|+\frac{t\lambda_+}{1+t}\bigg(|\Pi_1(B_n)|^2-|\Pi_1(B_n)|\bigg)\\\nonumber
 \geq &&-\frac{t\lambda_+}{(1+t)n_2 n_3}|G(B_n)|^2
\eeqa
Hence, with (\ref{50}) and (\ref{51}), we obtain that if (\ref{lambda+-}) holds and  $|G(B_n)|\geq n_1$,  
 \beqa
 0\leq &&\frac{t}{1+t}\frac{\lambda_+a^{-2}}{n_2 n_3}|G(B_n)|^2+\sum_{i,j\in\Pi_1(B_n)}(1-F_{i,j})v_{1,1}^a(x_i-x_j)\\\nonumber+&&\sum_{j\in\Pi_1( B_n), \,\,i\in\pi_2}(1-F_{i,j})[v^a_{2,1}]_-(x_i-x_j),
 \eeqa
Then summing up all the $B_n$'s satisfying $|G(B_n)|>n_1$, 
with (\ref{ij2}) and  $v_{11}\geq 0$, we obtain that as long as (\ref{lambda+-}) holds,
 \beqa\label{lambda+geq}
0\leq &&\sum_{i,j\in\pi_2}(1-F_{i,j})\half v^a_{2,2}(x_i-x_j)+\sum_{i,j\in\pi_1}(1-F_{i,j})\half v_{1,1}^a(x_i-x_j)\\\nonumber
 &&+\sum_{j\in\pi_1,i\in\pi_2}(1-F_{i,j})\half [v^a_{2,1}]_-(x_i-x_j)\big(1-\chi_A(x_j)\big)
 \eeqa
 Here $A$ is defined as  the set $\cup_{|G(B_n)|\leq n_1}B_n$.
 \beq\label{defA}
 A=\cup_{|G(B_n)|\leq n_1}B_n
 \eeq 
So far, we proved the interaction potential between particles of the same groups cancels out the negative part of  the $v^a_{2,1}(1-\chi_A)$ term in (\ref{sev21}). We shall show that the kinetic energy and the positive part of $v^a_{2,1}$ cancel out the remaining negative part of $v^a_{2,1}$.  
\par For the other terms in the Hamiltonian $H_P$, we claim that as long as 
\beq\label{riside0}
SL[4\,n_1(v+tv_-)]\geq 0
\eeq we have
\beqa
\label{riside}
0\leq &&\frac12\sum_{j\in\pi_1,i\in\pi_2}(1-F_{i,j})\bigg( [v^a_{2,1}]_+(x_i-x_j)+[v^a_{2,1}]_-(x_i-x_j)\chi_A(x_j)\bigg)\\\nonumber
&&+ \frac{2+t}{1+t}\sum_{j\in\pi_1}-\nabla_j\left(1-\sum_iF_{i,j}\right)\nabla_j
\eeqa
 As we can see that (\ref{lambda+geq}) and (\ref{riside}) implies that $H_P\geq 0$ when $SL[4n_1(v+tv_-)]\geq 0$ and   $(\ref{lambda+-})$ holds, i.e., $\lambda_+\geq \const(1+t^{-1})\lambda_-(R_0/r_1)^3$, which completes the proof of Theorem 2.
\par To prove (\ref{riside}), we only need to prove  the following operator inequality, for any fixed $x_2,\cdots, x_N$,
\beqa\label{last2}
0\leq&& -\frac{2+t}{1+t}\nabla_1(1-\sum_{i=2}^NF_{i,1})\nabla_1\\\nonumber&&+\half \sum_{j\in\pi_2}(1-F_{j,1})\bigg([v^a_{2,1}]_+(x_1-x_j)+[v^a_{2,1}]_-(x_1-x_j)\chi_A(x_1)\bigg)
\eeqa
First, if $[v^a_{2,1}]_-(x_1-x_j)\chi_A(x_1)\neq 0$, then $d(B^{x_1}_n,x_j)\leq R_0a$, here the $B^{x_1}_n$ is the cube where $x_1$ is. We obtain that $j\in \pi_2'\subset \pi_2$, here $\pi'_2$ is defined as 
\beq\label{defpi'2}
\pi'_2\equiv \{j\,'\in\pi_2:\exists B_n, D(x_{j\,'},B_n)\leq R_0a, |G(B_n)|\leq n_1\}
\eeq
Hence, it only remains to prove that
\beqa\label{last22}
0\leq&& -\frac{2+t}{1+t}\nabla_1(1-\sum_{i=2}^NF_{i,1})\nabla_1+\half\sum_{j\in \pi'_2}(1-F_{j,1}) v^a_{2,1}(x_1-x_j)
\eeqa
Second, we claim the following inequality which will be proved later. 
\beq\label{ei}
n_1\left(1-\sum_{i=2}^NF_{i,1}\right)\geq \sum_{j\in \pi'_2}(1-F_{j,1})\theta_{(R_0a)}(x_1-x_j)
\eeq
which implies that 
\beq\label{nabla1}
-\nabla_1\left(1-\sum_{i=2}^NF_{i,1}\right)\nabla_1\geq-\frac{1}{n_1}\nabla_1 \sum_{j\in \pi'_2}(1-F_{j,1})\theta_{(R_0a)}(x_1-x_j)\nabla_1
\eeq
With (\ref{nabla1}), we obtain that the right side of (\ref{last22}) is not less than
\beqa\nonumber
&&\sum_{j\in \pi'_2}\bigg(-\frac{2+t}{n_1(1+t)}\nabla_1(1-F_{j,1})\theta_{(R_0a)}(x_1-x_j)\nabla_1+\frac12(1-F_{j,1}) v^a_{2,1}(x_j-x_1)\bigg)\\\nonumber
\geq&&\sum_{j\in \pi'_2}\bigg(1-\prod_{k\neq 1\,or\,j}\bar \theta_{2R}(x_k-x_j)\bigg)\times \frac{2}{n_1(1+t)}\\\label{fin1}&&\times\bigg(-\nabla_1\theta_{(R_0a)}(x_1-x_j)\nabla_1+2n_1(v^a+tv^a_-)(x_j-x_1)\bigg)
\eeqa
Here we used the definition of $F_{j,1}$ and (\ref{defv1122}), i.e., $v^a_{2,1}=\frac{4}{1+t}[v^a+tv^a_-]$. With the assumption $SL[4\,n_1(v+v_-)]\geq 0$, we obtain that (\ref{fin1})$\geq 0$, which implies inequality (\ref{last22}).
\par Hence, it only remains to prove (\ref{ei}). For $x_2,\cdots,x_N$ fixed, we define $\pi_3$ as following, 
\beq
\pi_3=\left\{2\leq j\leq N: \prod_{2\leq k\leq N, k\neq j}\bar\theta_{2R}(x_j-x_k)=1\right\}
\eeq
With the definition of $\pi_3$, we obtain that
\begin{equation}
F_{j,1}=  \left\{ \begin{array}{ll} 
\theta_R(x_j-x_1) & j\in \pi_3 \\ 
0 & j\notin\pi_3\,,
\end{array}\right.
\end{equation}
Hence, it only remains to prove that
\beq
n_1\left(1-\sum_{i\in\pi_3}\theta_{R}(x_1-x_i)\right)\geq \sum_{j\in \pi'_2, j\notin\pi_3}\theta_{(R_0a)}(x_1-x_j)
\eeq
or 
\beq
\max_{x\in\R^3}\left(n_1\sum_{i\in\pi_3}\theta_{R}(x-x_i)+\sum_{j\in \pi'_2, j\notin\pi_3}\theta_{(R_0a)}(x-x_j)\right)\leq n_1
\eeq
Because the distance between $x_i$ ($i\in\pi_3$) and $x_j$ ($2\leq j\leq N, j\neq i$) are not less than $2R$, we have that if $i\in\pi_3$  
\beq
\theta_{R}(x-x_i)=1\Rightarrow\sum_{j\neq1, j\neq i}\theta_{R}(x-x_j)=0
\eeq
So, it only remains to prove that
\beq\label{n1geq}
\max_{x}\left(\sum_{j\in \pi'_2, j\notin\pi_3}\theta_{(R_0a)}(x-x_j)\right)\leq n_1
\eeq
By the definition of $\pi'_2$ in (\ref{defpi'2}), if $j\in \pi'_2$ and $\theta_{(R_0a)}(x-x_j)=1$, there exist $B_n$ satisfying $|G(B_n)|\leq n_1$ and $d(x, B_n)\leq 2R_0a$. Hence  by the definition of $G(B_n)$ in (\ref{defGb}) and (\ref{defGx}), we obtain that, for $\forall x\in\R^3$
\beq
\sum_{i\in \pi'_2}\theta_{(R_0a)}(x-x_i)=1\Rightarrow\sum_{i\in\pi_2}\theta_{(R_0a)}(x-x_i)\leq n_1
\eeq 
With the fact that $\pi'_2\subset \pi_2$, we arrive at the desired result (\ref{n1geq}) and complete the proof of Theorem 2. 
\end{proof}
\subsection{Proof of Lemma 1}
\begin{proof} Let $\delta \Omega$ be any infinitismal solid angle. With the definition of scattering length, we have that if $\phi$ is a complex-valued function such that 
\beq
\phi(x)=1\,\,\,{\rm for}\,\,\, x\in \delta\Omega\otimes\R \,\,\,{\rm and}\,\,\, |x|=R'\geq R_0a
\eeq
then
\beq
\delta \Omega \cdot a\leq \int_{\delta\Omega\otimes[0,R\,']}|\nabla\phi(x)|^2+\half v^a(x)|\phi(x)|^2dx
\eeq
Hence we obtain that
\beq
 a\int_{\R^3} \delta(|x|-R')|\phi(x)|^2dx\leq  \int_{|x|\leq R\,'}|\nabla\phi(x)|^2+\half v^a(x)|\phi(x)|^2dx
\eeq 
which says that for any non-negative radial function $U_0(x)$,
  supported in the annulus $R_0a\leq |x|\leq R$, with $\int_{\R^3}
  U_0(x)\,dx = 4\pi$, we have 
\beqa\label{115}
-\nabla\theta_R(x)\nabla+\half v^a\geq aU_0
\eeqa
\par Note: The result of lemma 2.5 of \cite{ERJJ2} shows the  $\theta_R(x)$ in above inequality can be replaced with the characteristic function of any star-shaped set when $v^a\geq 0$.
\par Furthermore, one can easily prove that for fixed $v$ ($SL[v]=1$), $v_+$ and small enough $\eps$, 
\beq
SL[v-\eps v_+]>0,\,\,\,\,\,\,|f_\eps|_{\infty}\leq \const
\eeq
Here we denote $f_\eps$ as the normalized solution ($\lim_{|x|\to \infty}f_\eps(x)=1$) of the zero-energy scattering equation of $v-\eps v_+$.
Hence, by the definition of scattering length, using $f_\eps$ as the trial function for $v$, we obtain that 
\beq
SL[v]\leq SL[v-\eps v_+]+\eps \|v_+\|_1 \cdot|f_\eps|_{\infty}\leq SL[v-\eps v_+]+\const\eps
\eeq
Combining this result with (\ref{115}) and the definition of $v^a_\eps$, we have, 
\beq
-(1-\eps)\nabla\theta_R(x)\nabla+\half v^a_\eps\geq (1-\const\eps)a U_0
\eeq
and 
\beq
-(1-\eps)\nabla_jF_{i,j}\nabla_j+\half F_{i,j}v^a_\eps(x_i-x_j)\geq (1-\const\eps)a F_{i,j}U_0(x_i-x_j)
\eeq
Hence, we obtain the following lower bound on  $H''$, which is defined in (\ref{lemone})
 \beqa
 H''\geq \eps\sum_{j}-\Delta_j+\eps \sum_{i\neq j}\frac{v^a_+}{2}(x_i-x_j)+(1-\const \eps)a\sum_{i\neq j}W_{i,j}
 \eeqa
Here  $W_{i,j}$ is defined as 
\beq
W_{i,j}=F_{ij}U_0(x_i-x_j)\geq 0
\eeq
As in \cite{ERJJ}, we choose 
\beq\ell=a Y^{-6}\eeq and divide $\Lambda$ into small cubes with side length $\ell$. Then we have 
\beq
H''\geq H^{(3)}\equiv\eps\sum_{j}-\Delta_j+\eps \sum_{i\neq j}\frac{v^a_+}{2}(x_i-x_j)+(1-\const \eps)a\sum_{i\neq j}W'_{i,j}
\eeq
Here $W'_{i,j}$ is defined as 
\beq
W'_{i,j}=G_{ij}U_0(x_i-x_j),\,\,\,\,G_{ij}\equiv F_{ij}\chi(x_i)\chi(x_j)\geq 0,
\eeq
and  $\chi(x)$ is equal to 1 when the distance between $x$ and the edges of the small cubes is greater than $2R$; otherwise it is equal to 0. As we can see the particles in different cubes don't affect each other in $H^{(3)}$.
\par We are going to estimate the ground energy $E^{(3)}(n,\ell)$ of $H^{(3)}$ for $n$ particles in $[0,\ell]^3$ with\textit{ Neumann} boundary condition. 
\par First, in the case that $n\leq\frac83\rho\ell^3Y^{-1}$,  with the definition of $\eps$ in (\ref{defeps}), we have that 
\beqa
H^{(3)}&&\geq 3Y\sum_{j}-\Delta_j+(1-\const Y)a\sum_{i\neq j}W'_{i,j}
\eeqa
Then with the Temple inequality in \cite{TE}, as in \cite{ERJJ2} (Ineq. 2.60, 2.66), we have that 
\beqa\nonumber
\frac{E^{\left(3\right)}\left(n,\ell\right)}{n}\geq &&4\pi  \frac{an}{\ell^3}\left(1-\frac1n\right)\left(1-\const Y\right)\left(1-\frac{2R}\ell\right)^3\left(1+\frac{4\pi n}{3}\left(\frac {2R}{\ell}\right)^3\right)^{-1}\\\label{126}
&&\left(1-\frac{3}{\pi}\frac{an}{\left(R^3-\left(aR_0\right)^3\right)\left(3\pi Y\ell^{-2}-4a\ell^{-3}n^2\right)}\right)\\\nonumber
\geq&&4\pi \frac{an}{\ell^3} (1-\frac1n)(1-\const Y)(1-\frac{n}{6\ell^3\rho}Y)
\eeqa
\par Second,  when $n\geq\frac83\rho\ell^3Y^{-1}$,  using the fact $W'\geq 0$, we obtain that
\beq
H^{(3)}\geq \eps H^{(4)}\equiv\eps\bigg(\sum_{j}-\Delta_j+ \sum_{i\neq j}\frac{v^a_+}{2}(x_i-x_j)\bigg)
\eeq
Using superadditivity of the ground state energy of $H^{(4)}$, we obtain that the ground energy $E^{(4)}(n,\ell)$  of $H^{(4)}$ is bounded from below as follows, ($n\geq p$)
\beqa
E^{(4)}(n,\ell)\geq \left[\frac n p\right]E^{(4)}(p,\ell)\geq \frac{n}{2p} E^{(4)}(p,\ell)
\eeqa
Here $[n/p\,]$ is the largest integer not greater than $n/p$. Actually, $H^{(4)}$ is just the Hamiltonian for the pure non-negative interaction potential, as in \cite{LY1}. Denote $a_+$ as follows:
\beq
a_+= \min \{SL(v^a), SL(v^a_+)\}\leq a
\eeq
Replacing $v_+^a$ with soft potential, we obtain that,
\beq
H^{(4)}\geq 3Y\sum_{j}-\Delta_j+(1-\const Y)a_+\sum_{i\neq j}W'_{i,j}
\eeq
As (\ref{126}), we can prove that when $p=\frac83\rho\ell^3Y^{-1}$,
 \beq
E^{(4)}(p,\ell)/p\geq \frac{32}3\pi a_+ \rho Y^{-1}\left(1-\frac12\right)(1-\const Y)\geq\frac{16}3\pi a_+ \rho Y^{-1}
\eeq
Hence when $n\geq\frac83\rho\ell^3Y^{-1}$, we have the following lower bound on  the ground energy $E^{(3)}(n,\ell)$ of $H^{(3)}$ 
\beq
E^{(3)}(n,\ell)/n\geq \eps E^{(4)}(n,\ell)/n\geq \frac{8}3\pi \eps a_+ \rho Y^{-1} \geq 8\pi a \rho
\eeq
For the last inequality, we used the definition of $\eps$ in (\ref{defeps}). So far we proved that 
\beqa\label{enell}
\frac{E^{(3)}(n,\ell)}{n}\geq \left\{ \begin{array}{ll} 
4\pi \frac{an}{\ell^3} (1-\frac1n)(1-\const Y)(1-\frac{n}{6\ell^3\rho}Y) & 1\leq n\leq \frac83\ell^3\rho Y^{-1} \\ 
8\pi a\rho & n\geq \frac83\ell^3\rho Y^{-1}\,,
\end{array}\right.
\eeqa
which implies that when $Y$ is small enough,
\beqa\label{enell2}
\frac{E^{(3)}(n,\ell)}{n(1-\const Y)}\geq \left\{ \begin{array}{ll} 
4\pi \frac{an}{\ell^3} (1-\frac1n) & 1\leq n\leq 4\ell^3\rho  \\ 
8\pi a\rho & n\geq 4\ell^3\rho\,,
\end{array}\right.
\eeqa
Recall the following two  facts, 
\begin{enumerate}
	\item the interaction potential $W'_{i,j}$ only depends on the particles in the same cubes as $i$ and $j$,
	\item the particles in different cubes have no interaction.
\end{enumerate}
Using the inequality above (\ref{enell2}), with the method in \cite{LY1} (Ineq. 2.55, 2.56), one can prove that the ground state $E^{(3)}(N,\Lambda)$ of $H^{(3)}$ of $N$ particles in big cubic $\Lambda$ is greater than
\beq
E^{(3)}(N,\Lambda)/N\geq 4\pi a\rho(1-\const Y).
\eeq 
Here $Y$ is defined in (\ref{defY}), which implies the desired result (\ref{lemone}).
\end{proof}

\begin{section} {Appendix}
In this appendix, we show that if  $v^a$ is a continuous function and $H_N$ has no bound state for any $N$, $v^a$ has a positive core and bounded from below, i.e., 
\beq v^a(0)>0,\,\,\,\,\min v^a(r)\neq -\infty\eeq 
And these inequalities also hold when $v^a$ is stable \cite{DavidR} in the sense of (\ref{appendix1}).
One can see that $\min v^a(r)\neq -\infty$ is trivial when $v^a$ is continuous. So it only remains to prove that $v^a(0)>0$.
\par First, we prove the statement in the case when $v^a$ is stable, which is defined as follows: there exists constant $C$, for any $N$, $x_1,\cdots,x_N$,
\beq\label{appendix1}
\sum_{1\leq i\neq j\leq N}v^a(x_i-x_j)\geq -CN,
\eeq
Inserting  
\beq x_1=x_2=\cdots=x_{[N/2]}=0,\,\,\,\,x_{[N/2]+1}=x_{[N/2]+2}=\cdots=x_{N}=x_0
\eeq
into the left side of (\ref{appendix1}), for some $x_0\in\R^3$ satisfying $v^a(x_0)<0$, we obtain that 
\beq\label{32}
\const v^a(0) N^2-\const v^a(x_0)N^2\geq -CN ,
\eeq
which implies the desired result that $v^a(0)>0$.
\par Next, we prove the statement in the case that $H_N$ has no bounded state for any $N$. Because $v^a$ is not pure non-negative, there exist $x_0\in\R^3,r_1,C\in R$ satisfying that
\beq
v^a(x)<-C,\,\,\,{\rm for}\,\,\,x\in B(x_0,r_1)\subset \R^3
\eeq 
Here $B(x_0,r_1)$ is the sphere of radius $r_0$ centered at $x_0$. If $v^a(0)\leq0$, there exists $r_2<r_1/2$ satisfying that \beq
v^a(x)<C/2,\,\,\,{\rm for}\,\,\, x\in B(0, r_2)
\eeq
We construct the trial state such that $x_1,x_2,\cdots,x_{[N/2]}$ are localized in $B(0,r_2)$ with the Dirichelet boundary condition and $x_{[N/2]+1},x_{[N/2]+2},\cdots,x_{N}$ are localized in $B(x_0,r_2)$ with the same boundary condition. The energy of this state is less than 
\beq
-\frac C8N^2 +\frac{\const}{r_2^2} N
\eeq
Here the first term is potential energy and the second term is kinetic energy. When $N$ goes to infinity, the energy of this trial state is negative and hence there are bound states, which is a contradiction with our assumptions. So we arrive at the desired result that $v^a(0)>0$.
\par
\end{section}

\end{document}